\documentclass[12pt]{article}

\usepackage{authblk}
\usepackage{geometry}
\usepackage{amsmath,amssymb,mathrsfs}
\usepackage{color}
\usepackage[dvipsnames]{xcolor}
\usepackage{graphicx}
\usepackage{diagbox}
\usepackage{soul}
\usepackage{algorithm,algpseudocode}
\usepackage{lineno,hyperref}
\usepackage{subfigure}
\usepackage{makecell}
\usepackage{booktabs}
\usepackage{stmaryrd}

\SetSymbolFont{stmry}{bold}{U}{stmry}{m}{n}
\usepackage{bm}
\SetSymbolFont{stmry}{bold}{U}{stmry}{m}{n}
\modulolinenumbers[5]

\DeclareMathOperator*{\Loss}{Loss}

\def\bx{\mathbf{x}}

\def\bn{\mathbf{n}}

\def\bft{\bm{\theta}}

\def\pd{\partial}
\newcommand{\beq}{\begin{equation}}
\newcommand{\eeq}{\end{equation}}
\newcommand{\beqs}{\begin{eqnarray}}
\newcommand{\eeqs}{\end{eqnarray}}
\newcommand{\beqsn}{\begin{eqnarray*}}
\newcommand{\eeqsn}{\end{eqnarray*}}
\newcommand{\bary}{\begin{array}}
\newcommand{\eary}{\end{array}}
\newcommand{\fr}[2]{\frac{#1}{#2}}

\newcommand*{\dbblk}[1]{\llbracket #1 \rrbracket}
\newcommand*{\dbblkL}[1]{{\biggl\llbracket} {#1} {\biggr\rrbracket}}

\title{Physics-Informed Machine Learning for Two-Phase Moving-Interface and Stefan Problems}
\author[1]{Che-Chia Chang}
\author[2,3]{Te-Sheng Lin}
\author[2]{Ming-Chih Lai}

\affil[1]{Institute of Artificial Intelligence Innovation, National Yang Ming Chiao Tung University, Hsinchu 30010, Taiwan}
\affil[2]{Department of Applied Mathematics, National Yang Ming Chiao Tung University, Hsinchu 30010, Taiwan}
\affil[3]{National Center for Theoretical Sciences, National Taiwan University, Taipei 10617, Taiwan}

\begin{document}

\maketitle

\begin{abstract}
The Stefan problem is a classical free-boundary problem that models phase-change processes and poses computational challenges due to its moving interface and nonlinear temperature-phase coupling. In this work, we develop a physics-informed neural network framework for solving two-phase Stefan problems. The proposed method explicitly tracks the interface motion and enforces the discontinuity in the temperature gradient across the interface while maintaining global consistency of the temperature field. Our approach employs two neural networks: one representing the moving interface and the other for the temperature field. The interface network allows rapid categorization of thermal diffusivity in the spatial domain, which is a crucial step for selecting training points for the temperature network. The temperature network's input is augmented with a modified zero-level set function to accurately capture the jump in its normal derivative across the interface. Numerical experiments on two-phase dynamical Stefan problems demonstrate the superior accuracy and effectiveness of our proposed method compared with the ones obtained by other neural network methodology in literature. The results indicate that the proposed framework offers a robust and flexible alternative to traditional numerical methods for solving phase-change problems governed by moving boundaries. In addition, the proposed method can capture an unstable interface evolution associated with the Mullins-Sekerka instability.
\end{abstract}

\section{Introduction}
\label{sec:intro}

Moving interface problems arise in a wide range of physical and engineering applications, where the evolution of an interface between distinct phases plays a central role. Such problems typically involve partial differential equations (PDEs) defined in subdomains separated by a moving interface, whose motion is governed by interfacial conditions derived from physical laws. Examples include multiphase fluid flow, crystal growth, solidification, and other phase-change processes. Among these, the Stefan problem---named after Josef Stefan, who introduced it in 1891~\cite{Stefan1891}---serves as a prototypical model describing heat transfer with phase change, where the temporal evolution of an interface separates different material phases, such as solid and liquid. In the Stefan problem, each phase satisfies a heat equation, while the interface motion is governed by an energy balance that accounts for latent heat effects. This problem not only captures essential physical mechanisms in melting and solidification but also exemplifies the mathematical and numerical challenges inherent in solving free boundary problems.

Analytical solutions to the Stefan problem exist only for idealized configurations, typically in one-dimensional or self-similar geometries~\cite{Crank87, Gupta17}. For general multidimensional cases, various numerical methods have been developed, including front-tracking schemes~\cite{Womble90, UT92, JT96}, enthalpy~\cite{Voller81, Voller87, Date92}, and phase-field formulations~\cite{Fix82, MR02}, and level-set approaches~\cite{OS88, GF05, PHRG13}. Despite substantial progress, accurately capturing the interface motion and ensuring consistency in the heat flux across discontinuities remain challenging, especially in higher-dimensional problems with strong nonlinearities or discontinuous thermal properties~\cite{Furzeland80, JVV06}.

In recent years, machine learning techniques, particularly neural networks, have emerged as a promising alternative framework for solving PDEs. Physics-informed neural networks~(PINNs)~\cite{RPK19} integrate data-driven learning and physics-based modeling by leveraging automatic differentiation to enforce governing equations and boundary conditions during training. Despite their success, standard PINN formulations often struggle to capture discontinuities or sharp gradients in the solution or its derivatives unless special treatments are introduced in the neural network architecture~\cite{HLL22} or the loss function~\cite{LCLHL22}.

To address such challenges in free-boundary problems, Wang et al.~\cite{WP21} proposed a deep learning-based Stefan problem solver employing two neural networks---one for the temperature field and another for the interface---as hypothesis functions for learning. Another approach, the deep level set method~\cite{SS24}, represents the moving solid-liquid interface using a neural network-parameterized level-set function. Lin et al.~\cite{LZZZ25} extended the discontinuous extreme learning machine framework to free-boundary problems by incorporating a discrete time-stepping scheme to evolve the interface and solve the PDE sequentially.

In this study, we employ the PINNs framework to investigate the two-dimensional, two-phase Stefan problem, characterized by a discontinuous thermal diffusivity across the moving interface. To accurately resolve the interfacial discontinuity in the temperature gradient, we adopt a cusp-capturing PINN~\cite{TLHL23} formulation specifically designed to handle gradient discontinuities. The proposed approach is validated through a series of numerical experiments that demonstrate its accuracy and robustness in solving the two-phase Stefan problem. The remainder of this paper is organized as follows. Section~2 presents the mathematical formulation of the problem. Section~3 introduces the PINN methodology. Numerical results are discussed in Section~4, and concluding remarks are provided in Section~5.

\section{Two-dimensional two-phase Stefan (dynamical interface) problem}
\label{sec:stefan}

Consider a two-dimensional square domain $\Omega = (0, x_L) \times (0, y_L)$, which is divided by a moving interface $\Gamma(t)$ defined as
\begin{equation}
    \Gamma(t) = \{(x, y)\in\Omega \,|\, x=s(y,t), \, y\in(0, y_L)\}.
\end{equation}
Accordingly, the domain can be decomposed as
$$
    \Omega = \Omega^+(t) \cup \Omega^-(t) \cup \Gamma(t),
$$
where $\Omega^+(t)=\{ (x,y,t) \mid x>s(y,t)\}$ and $\Omega^-(t)=\{ (x,y,t) \mid x < s(y,t)\}$. Let $u(x,y,t)$ denote the temperature function that is a continuous function and satisfies the heat equation in each subdomain, characterized by thermal diffusivities $k^+$ and $k^-$, respectively. Specifically,
\begin{equation}
    \frac{\partial u}{\partial t}  =  k^{\pm} \Delta u, \quad \text{in} \quad \Omega^{\pm}, \label{heat}
\end{equation}
where $\Delta u = \partial^2_x u + \partial^2_y u$. We impose a Dirichlet boundary condition at $x=0$ and Neumann boundary conditions on the remaining boundaries:
\begin{eqnarray}
    u(0,y,t) &=& -|U_0|, \quad y\in (0, y_L), \\
    \frac{\pd u}{\pd \bn} &=&  0, \quad \text{on} \quad \pd \Omega \setminus \{x=0\},  \label{insulated}
\end{eqnarray}
with given $U_0(y,t)$. The initial temperature is prescribed as
\begin{equation}
    u(x, y, 0)  =  u_0(x, y), \quad \text{in} \quad \Omega. \label{init}
\end{equation}

The moving interface $\Gamma(t)$ between the two phases is not known \emph{a priori} and is determined as part of the solution. The melting temperature is assumed to be zero along the interface, yielding
\begin{equation}
    u(x, y, t)  = 0,   \quad   \text{on}  \quad \Gamma(t).  \label{freezing}
\end{equation}
The motion of the interface is governed by the Stefan condition \cite{Crank87}, which relates the normal velocity of the moving boundary to the discontinuity in heat flux across the interface:
\begin{equation}
    \beta \, V_n(x, y, t)  =  - \dbblkL{k \frac{\partial u}{\partial \bn}} (x, y, t),    \quad    \text{on}  \quad \Gamma(t),
    \label{moving}
\end{equation}
where $\beta$ is the Stefan number, $\dbblk{\cdot}$ represents the quantity jump across $\Gamma(t)$ from $\Omega^+$ to $\Omega^-$, and $V_n$ denotes the normal velocity of the interface. By introducing the level set function $\phi(x, y, t) = x - s(y,t)$, the normal velocity can be expressed as
\beq
V_n(x,y,t) = - \frac{\phi_t(x,y,t)}{\|\nabla \phi(x,y,t) \|}\quad    \text{on}  \quad \Gamma(t). \label{vn}
\eeq
Consequently, by substituting Eq.~(\ref{vn}) into Eq.~(\ref{moving}), we obtain the evolution equation for the interface position $s(y,t)$:
\beq
\left(\frac{\beta}{\sqrt{1+s^2_y}}\right)\, \fr{\pd s}{\pd t} = - \dbblkL{k \frac{\partial u}{\partial \bn}} (s(y,t), y, t), \quad y\in(0, y_L), \quad t>0. \label{moving2}
\eeq
Finally, the initial position of the interface is specified as
\begin{equation}
    s(y, 0) =  s_0(y),    \quad                                                                                                                                      y \in (0, y_L). \label{init2}
\end{equation}

Notice that the jump quantity in the right-hand side of Eq.~(\ref{moving2}) can be written as
\begin{align}
    \dbblkL{k \frac{\partial u}{\partial \bn}}
     & =k^+ \frac{\partial u^+}{\partial \bn}  - k^- \frac{\partial u^-}{\partial \bn}  \nonumber                                                                                                                                                                                       \\
     & =k^+ \frac{\partial u^+}{\partial \bn}  + k^- \frac{\partial u^+}{\partial \bn}  - k^- \frac{\partial u^+}{\partial \bn}  - k^- \frac{\partial u^-}{\partial \bn}                                                                                                      \nonumber \\
     & = \dbblk{k} \frac{\partial u^+}{\partial \bn} + k^-\dbblkL{\frac{\partial u}{\partial \bn}}. \label{moving3}
\end{align}
So, even when the thermal diffusivity is continuous, i.e., $\dbblk{k}=0$, the jump of the heat flux $\dbblk{k \frac{\partial u}{\partial \bn}}$ remains nonzero as long as the interface is moving. Consequently, the temperature $u$ is continuous across the interface, whereas its normal derivative exhibits a jump discontinuity at the interface. In this work, we focus on solving the Stefan problem described by Eqs.~(\ref{heat})-(\ref{freezing}), (\ref{moving2}), and (\ref{init2}) using physics-informed neural networks~\cite{RPK19}.

\section{Solving the Stefan problem using cusp-capturing PINNs}
\label{sec:stefanPINN}

Since the temperature \(u\) is continuous across the interface while its normal derivative exhibits a jump discontinuity, it is natural to employ the cusp-capturing PINNs developed by Tseng et al.~\cite{TLHL23} to address this problem. In this framework, a modified level-set function is introduced as an additional input to the neural network, enabling it to accurately capture solutions of PDEs with cusp singularities.

Specifically, we denote the interface hypothesis function by $\hat{s}(y, t; \theta_2)$, which is modeled using a neural network with a smooth activation function, and $\theta_2$ is the set of trainable parameters associated with the interface network; consequently, the interface can be expressed as
\begin{equation}
    \Gamma(t)=\{(x, y) | x = \hat{s}(y, t; \theta_2), \quad y\in(0, y_L), \quad t\ge 0\}.
\end{equation}
The level set function and its modified counterpart are then defined as \(\phi(x,y,t; \theta_2)=x-\hat{s}(y,t; \theta_2)\) and \(\phi^a(x,y,t; \theta_2)=|x-\hat{s}(y,t; \theta_2)|\), respectively. The hypothesis function for the temperature is given by
\begin{equation}
    \hat{u}(\bx, t; \bft) = U(\bx, t, \phi^a(\bx, t; \theta_2); \theta_1), \label{net-u}
\end{equation}
where \(\bx = (x, y)\), $U=U(\bx, t, z; \theta_1)$ denotes the neural network function with a smooth activation function and $\theta_1$ represents the set of its trainable parameters, while $\bft$ is the collection of all the trainable parameters that is the union of $\theta_1$ and $\theta_2$.

The derivatives of \(\hat{u}\) in $\Omega^\pm$ can be calculated using chain rule as
\begin{equation}
    \nabla_{\bx} \hat{u} = \nabla_{\bx} U + \partial_z U \nabla_{\bx} \phi^a, \label{grad}
\end{equation}
where \(\nabla_{\bx}\) is the vector of the partial derivatives with respect to \(\bx\), and \(\partial_z U\) is the partial derivative of \(U\) with respect to \(z\). We can immediately see that the normal derivative of $\hat{u}$ has a jump discontinuity at the interface $x=\hat{s}(y,t; \theta_2)$ when using the modified level set function $\phi^a(\bx, t; \theta_2)$. This behavior is essential for accurately modeling the Stefan problem. Then, using the fact that normal vector at the interface \(\bn = \nabla_{\bx} \phi/\|\nabla_{\bx} \phi\| = (1, -\partial_y \hat{s})^T/\sqrt{1+(\partial_y \hat{s})^2} \), we can calculate the flux jump at the interface as
\begin{align}
    \dbblkL{k \frac{\partial \hat{u}}{\partial \bn}} & = \dbblkL{k \nabla_{\bx} U \cdot \bn} + \dbblkL{k \partial_z U \nabla_{\bx} \phi^a \cdot \bn} \nonumber                                            \\
                                                     & = (k^+ - k^-) \nabla_{\bx} U \cdot \bn + k^+ \partial_z U \nabla_{\bx} \phi \cdot \bn - k^- \partial_z U (-\nabla_{\bx} \phi) \cdot \bn \nonumber  \\
                                                     & = (k^+ - k^-) \nabla_{\bx} U \cdot \bn + (k^+ + k^-) \partial_z U \nabla_{\bx} \phi \cdot \bn \nonumber                                            \\
                                                     & = (k^+- k^-) \nabla_{\bx} U \cdot \frac{\nabla_{\bx} \phi}{\|\nabla_{\bx} \phi\|} + (k^+ + k^-) \partial_z U \|\nabla_{\bx} \phi\|. \label{normal}
\end{align}

It is worth noting that, unlike the cusp-capturing PINNs presented in \cite{TLHL23}, the interface $\hat{s}(y, t; \theta_2)$ in this study is not known a priori but rather determined as part of the solution. With this in mind, we now proceed to define the loss function for the Stefan problem.

\subsection{Loss function}

We sample \(M_r\) points from the interior domain,
$$
    \{(\bx^i, t^i)\}_{i=1}^{M_r} = \{(x^i, y^i, t^i)\}_{i=1}^{M_r},\quad \bx^i\in\Omega^+(t)\cup\Omega^-(t), \quad t^i\in (0, T],
$$
\(M_0\) points for the initial condition,
$$
    \{(\bx^i_0, 0)\}_{i=1}^{M_0} = \{(x_0^i, y_0^i, 0)\}_{i=1}^{M_0}, \quad \bx^i_0\in\Omega,
$$
\(M_b\) points for the boundary conditions,
$$
    \{(\bx^i_b, t^i_b)\}_{i=1}^{M_b} = \{(x_b^i, y_b^i, t_b^i)\}_{i=1}^{M_b}, \quad \bx^i_b\in\partial\Omega, \quad t^i_b\in(0, T],
$$
and \(M_I\) points for the interface,
$$
    \{(\bx^i_I, t^i_I)\}_{i=1}^{M_I}, \quad \bx^i_I\in\Gamma(t), \quad t^i_I\in(0, T].
$$
For the interface sampling, we first drawn \(M_I\) points, $\{(y^i_I, t^i_I)\}^{M_I}_{i=1}\subset (0, y_L) \times (0, T)$, and each interface point \((\bx^i_I, t^i_I)\) is then defined as \((\hat{s}(y^i_I, t^i_I; \theta_2), y^i_I, t^i_I)\). With these sampling procedures established, the loss function can be formulated accordingly as
\begin{align}
    \begin{split}
        \text{Loss}(\bft)
        & = \frac{1}{M_r} \sum_{i=1}^{M_r}\left|L_r(\bx^i, t^i; \bft)\right|^2
        + \frac{1}{M_0} \sum_{i=1}^{M_0}\left|L_0(\bx^i_0, 0; \bft)\right|^2 \\
        & + \frac{1}{M_b} \sum_{i=1}^{M_b}\left|L_{b}(\bx^i_{b}, t^i_b; \bft)\right|^2
        + \frac{1}{M_I} \sum_{k=1}^3 \sum_{i=1}^{M_I} \left|L_{I_k}(\bx^i_{I}, t^i_I; \bft)\right|^2,
    \end{split} \label{loss}
\end{align}
where the individual loss components are defined as
\begin{align}
    L_r(\bx^i, t^i; \bft)         & = \frac{\partial \hat{u}}{\partial t}(\bx^i, t^i; \bft) - k^i \Delta \hat{u}(\bx^i, t^i; \bft),                          \label{loss:interior} \\
    L_0(\bx^i_0, 0; \bft)         & =  \hat{u}(\bx^i_0, 0; \bft) - u_0(\bx^i_0) ,                                                                                                  \\
    L_b(\bx_b^i, t_b^i; \bft)     & =
    \begin{cases}
        \hat{u}(\bx_b^i, t_b^i; \bft) + |U_0(\bx_b^i, t_b^i)|,     & \text{if } x_b^i = 0,                          \\
        \frac{\partial \hat{u}}{\partial n}(\bx_b^i, t_b^i; \bft), & \text{if } x_b^i \neq 0, \label{loss:boundary}
    \end{cases}                                                     \\
    L_{I_1}(\bx_I^i, t^i_I; \bft) & = \hat{u}(\bx_I^i, t^i_I; \bft) ,                                                                                                              \\
    L_{I_2}(\bx_I^i, t^i_I; \bft) & = \beta \, V_n(\bx_I^i, t^i_I) + \dbblkL{k \frac{\partial \hat{u}}{\partial \bn}}(\bx_I^i, t^i_I; \bft), \label{loss:interface}                \\
    L_{I_3}(\bx_I^i, 0; \bft)     & = \hat{s}(y_I^i, 0; \theta_2) - s_0(y_I^i),
\end{align}
where \(k^i\) is the thermal diffusivity at the point \(\bx^i\).

There are several important remarks to be made here. First, the loss function follows the standard physics-informed neural network framework, in which it is defined as the mean-squared error of the residuals corresponding to the governing PDE, the initial condition, the boundary condition, and the interface condition. Second, following the definition of \(u_{\bft}\) in Eq.~(\ref{net-u}), the computation of its derivative requires additional care, since the network involves an auxiliary variable \(z=\phi^a = |x- \hat{s}(y, t; \theta_2)|\). Using the automatic differentiation framework in PyTorch, the derivatives appear in Eqs.~(\ref{loss:interior}) and (\ref{loss:boundary}) can be evaluated directly. Meanwhile, the right-hand side of Eq.~(\ref{loss:interface}) can be written explicitly as
\begin{align}
    \begin{split}
        & \beta \, V_n(\bx_I^i, t^i_I; \bft) + \dbblkL{k \frac{\partial \hat{u}}{\partial \bn}}(\bx_I^i, t^i_I; \bft)  \\
        =& \frac{\beta}{\sqrt{1 + \left(\partial_y \hat{s}(y^i_I,t^i_I; \theta_2)\right)^2}}  \frac{\partial}{\partial t}\hat{s}(y^i_I,t^i_I; \theta_2) \\
        &+ (k^+ - k^-) \nabla_{\bx} U(\hat{s}(y^i_I, t^i_I; \theta_2), y^i_I, t^i_I; \theta_1) \cdot \frac{\left(1, -\partial_y \hat{s}(y^i_I,t^i_I; \theta_2)\right)^T}{\sqrt{1 + \left(\partial_y\hat{s}(y^i_I,t^i_I; \theta_2)\right)^2}} \\
        &+(k^- + k^+) \partial_zU(\hat{s}(y^i_I,t^i_I; \theta_2), y^i_I, t^i_I; \theta_1) \sqrt{1 + \left(\partial_y \hat{s}(y^i_I,t^i_I; \theta_2)\right)^2}. \label{loss_snc}
    \end{split}
\end{align}
All the above derivatives can be computed directly using automatic differentiation. Finally, in Eq.~(\ref{loss:interior}), the thermal diffusivity \(k^i\) at each point \(\bx^i\) is determined by identifying whether the point lies in \(\Omega^+(t)\) or \(\Omega^-(t)\), and assigning \(k^i = k^+\) or \(k^-\), respectively. This can be done by comparing the \(x\) coordinate of the point with  \(\hat{s}(y^i, t^i; \theta_2)\). That is, for each point \(\bx^i\), we set \(k^i = k^+\) if \(x^i > \hat{s}(y^i, t^i; \theta_2)\) and \(k^i = k^-\) if \(x^i < \hat{s}(y^i, t^i; \theta_2)\). This categorization step is performed whenever the interface is updated.

To minimize the loss function \(\text{Loss}(\bft)\), we adopt the Levenberg-Marquardt~(LM)~\cite{ST12} algorithm to update the parameters \(\bft\). The LM algorithm is a classical optimization method that blends the advantages of gradient descent and the Gauss-Newton method, providing stable progress far from the optimum and faster convergence near it. In each iteration, the parameter increment $\Delta\bft$ is computed and the parameters are updated as $\bft \gets \bft + \Delta \bft$. The update is accepted when it results in a decrease in the loss; otherwise, it is rejected. 

The LM algorithm has been shown to be effective for training PINNs~\cite{TLHL23}, and this training procedure can be readily modified to other optimization problems that rely on iterative parameters update. We summarize the overall training procedure in Algorithm~\ref{alg:train}, and the categorization of the thermal diffusivity is described in Algorithm~\ref{alg:track}. One thing to note is that Algorithm~\ref{alg:track} can be efficiently implemented in PyTorch using the \texttt{torch.where} function, thereby avoiding explicit for-loops and improving computational efficiency during training. Additionally, the update and rejection rules in Algorithm~\ref{alg:train} can be further optimized depending on the specific training task. For brevity, we omit these details here.

\begin{algorithm}[!ht]
    \caption{Training process}\label{alg:train}
    \begin{algorithmic}
        \State Sample the training points \(\{(\bx^i, t^i)\}_{i=1}^{M_r}\), \(\{(\bx^i_0, 0)\}_{i=1}^{M_0}\), \(\{(\bx^i_b, t^i_b)\}_{i=1}^{M_b}\), \(\{(\bx^i_I, t^i_i)\}_{i=1}^{M_I}\)
        \State  \(\{k^i\}_{i=1}^{M_r}\), \( \gets \) {\textsc{CategorizeThermalDiffusivity}($\bft$)}
        \For {\(n=1,2,\cdots,N\)}
        \State Calculate \(\text{Loss}(\bft)\)
        \State Calculate potential update \(\Delta \bft\) (e.g., using LM optimizer)
        \State \(\bft \gets \bft + \Delta \bft\)
        \State \(\{k^i\}_{i=1}^{M_r}\), \(\gets \) {\textsc{CategorizeThermalDiffusivity}($\bft$)}
        \State Calculate updated \(\text{Loss}(\bft)\)
        \If{update is rejected (e.g., Loss value increases)}
        \State \(\bft \gets \bft - \Delta \bft\)
        \State \(\{k^i\}_{i=1}^{M_r}\), \( \gets \) {\textsc{CategorizeThermalDiffusivity}($\bft$)}
        \EndIf
        \EndFor
    \end{algorithmic}
\end{algorithm}
\begin{algorithm}[!ht]
    \caption{Categorize thermal diffusivity}\label{alg:track}
    \begin{algorithmic}
        \Function{CategorizeThermalDiffusivity}{$\bft$}
        \For {\(i=1,2,\cdots,M_r\)}
        \If{\(x^i_r < \hat{s}(y^i_r, t^i_r; \theta_2)\)}
        \State \(k^i \gets k^-\)
        \Else
        \State \(k^i \gets k^+\)
        \EndIf
        \EndFor
        \State \Return  \(\{k^i\}_{i=1}^{M_r}\)
        \EndFunction
    \end{algorithmic}
\end{algorithm}

\section{Numerical results}
\label{sec:numerics}

In this section, we present several numerical experiments to demonstrate the effectiveness of our proposed methodology for solving the Stefan problem. Throughout all the numerical experiments, two fully connected feed-forward neural networks with Sigmoid activation function are employed for \(U_{\theta_1}\) and \(\hat{s}_{\theta_2}\), respectively, which are coupled through the loss function given in Eq.~(\ref{loss}). The architecture of each network is denoted by \((d_{\text{in}}, n_1, n_2, \ldots, 1)\), where \(d_{\text{in}}\) represents the input dimension, \(n_1, n_2, \ldots\) denote the number of neurons in each hidden layer, and $1$ is the output dimension.

As stated in the previous section, the parameters \(\bft\) are updated using the LM optimizer, and the training points are generated via the Latin hypercube sampling algorithm, ensuring a relatively uniform distribution of samples. The training process is terminated once the loss function decreases below \(10^{-16}\).

We denote \(\tilde{M}_b = M_0 + M_b + 3 M_I\) as the total number of points associated with the initial, boundary, and interface conditions. Hence, the total number of training points is given by \(M = M_r + \tilde{M}_b\). For some of the examples, a heat source term is added to the governing equation so that the prescribed solution exactly satisfies the heat equation with the source term.

\paragraph{Example 1-1}

In the first test, we aim to demonstrate the effectiveness and accuracy of the present PINN approach for solving the Stefan problem and to compare its performance with the deep learning method developed in \cite{WP21}. The test example, taken from Section~3.2.1 of \cite{WP21}, corresponds to a one-dimensional two-phase Stefan problem. Although the algorithm presented earlier is formulated for two-dimensional two-phase problems, the proposed methodology certainly can be applied to one-dimensional cases with even greater simplicity.

Let us set \(\Omega = (0, 2)\), \(t \in (0, 1)\), \(k^- = 1\), \(k^+ = 2\), and \(\beta = 1\). The exact solution is given by
\[
    u(x,t) =
    \begin{cases}
        \exp\left(\frac{1}{2} + t - x\right) - 1,                   & x \leq  t + \frac{1}{2}, \\
        2\left(\exp\left((\frac{1}{2} + t - x)/2\right) - 1\right), & x > t + \frac{1}{2}.
    \end{cases}
\]
The initial condition is prescribed directly from the exact solution, and Dirichlet boundary conditions are imposed along the domain boundaries. It can be readily verified that both the melting temperature condition Eq.~(\ref{freezing}) and the Stefan condition Eq.~(\ref{moving}) are satisfied at the moving interface $x=s(t) = t+1/2$.

We train the network using the LM optimizer for a maximum of $2000$ iterations. For comparison, we also implemented the deep learning method from \cite{WP21} and trained the networks using the same LM optimizer and number of iterations reported in that work. In Table \ref{tab:ex1_errors}, we show the relative \(L^2\) errors of \(\hat{u}\) and \(\hat{s}\) for different methods and network architectures. The errors are computed by sampling \(M_r=10^6\) and \(\tilde{M}_b=8\times 10^6\) random points, averaged over ten independent runs. In each run, the training points are sampled randomly, and the network parameters are initialized using PyTorch default settings. As shown in the table, even with a network consisting of a single hidden layer with $32$ neurons, our method can achieve a relative \(L^2\) error on the order of $10^{-9}$ for both \(u\) and \(s\). The errors obtained by the present method are significantly smaller ($O(10^{-9})$ versus $O(10^{-4})$) than those reported for the deep learning method in \cite{WP21}, which employs a three-hidden-layer network with $100$ neurons in each hidden layer. Moreover, when using the same training algorithm as in \cite{WP21} but replacing the Adam optimizer with the LM optimizer, the temperature error can be reduced to $O(10^{-8})$, while the interface error remains at $O(10^{-4})$. These results show that the present network methodology substantially outperforms the algorithm in \cite{WP21}.

\begin{table}[!ht]
    \begin{tabular}{c|c|cc}
        \toprule
        Method                   & Network Architectures              & $\|\hat{u}-u\|_2/\|u\|_2$ & $\|\hat{s}-s\|_2/\|s\|_2$ \\
        \hline
        Wang et al. \cite{WP21}  & \makecell{u: (2, 100, 100, 100, 2)                                                         \\ s: (1, 100, 100, 100, 1)} & 5.83e-04                         & 2.81e-04                                 \\
        \hline
        Wang et al. \cite{WP21}* & \makecell{u: (2, 100, 2)                                                                   \\ s: (1, 100, 1)}                     & 3.63e-08                     & 2.35e-04               \\
        \hline
        Present                  & \makecell{u: (3, 32, 1)                                                                    \\ s: (1, 32, 1)}             & 2.05e-09                      & 5.65e-09              \\
        \bottomrule
    \end{tabular}
    \caption{Relative \(L^2\) errors of \(\hat{u}\) and \(\hat{s}\) for Example 1-1. The asterisk (*) indicates that we implemented the method proposed in \cite{WP21} and trained the networks using the LM optimizer instead of the original training strategy. The errors are averaged across ten runs. The values reported in Wang et al.~\cite{WP21} are taken directly from the paper.}
    \label{tab:ex1_errors}
\end{table}

\begin{figure}[!ht]
    \centering
    \subfigure[]{\label{fig:ex1_a}\includegraphics[width=\textwidth]{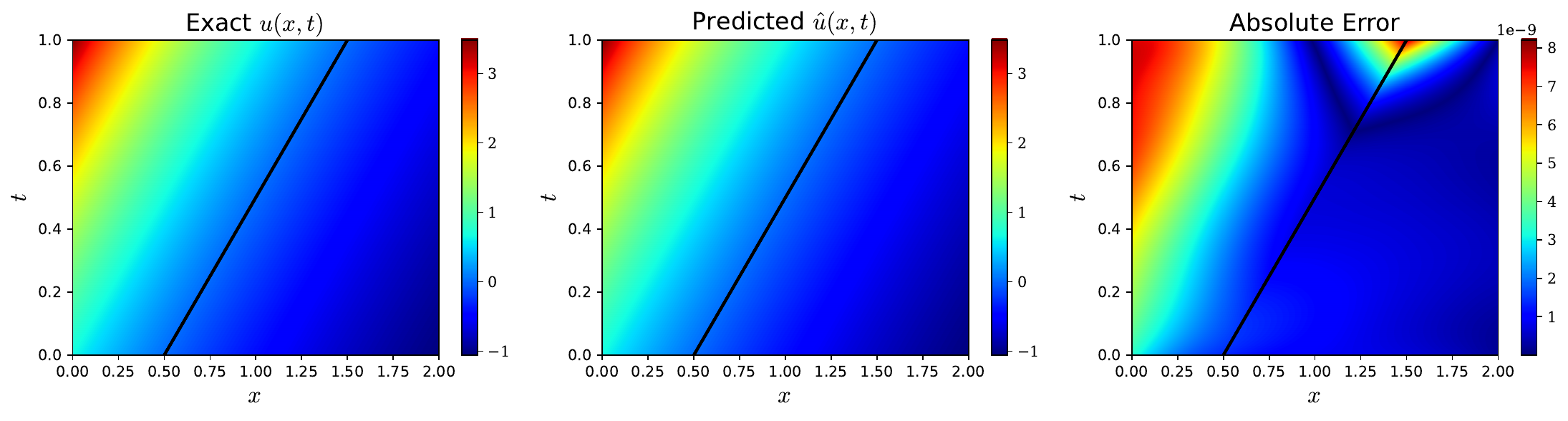}}
    \subfigure[]{\label{fig:ex1_b}\includegraphics[width=.75\textwidth]{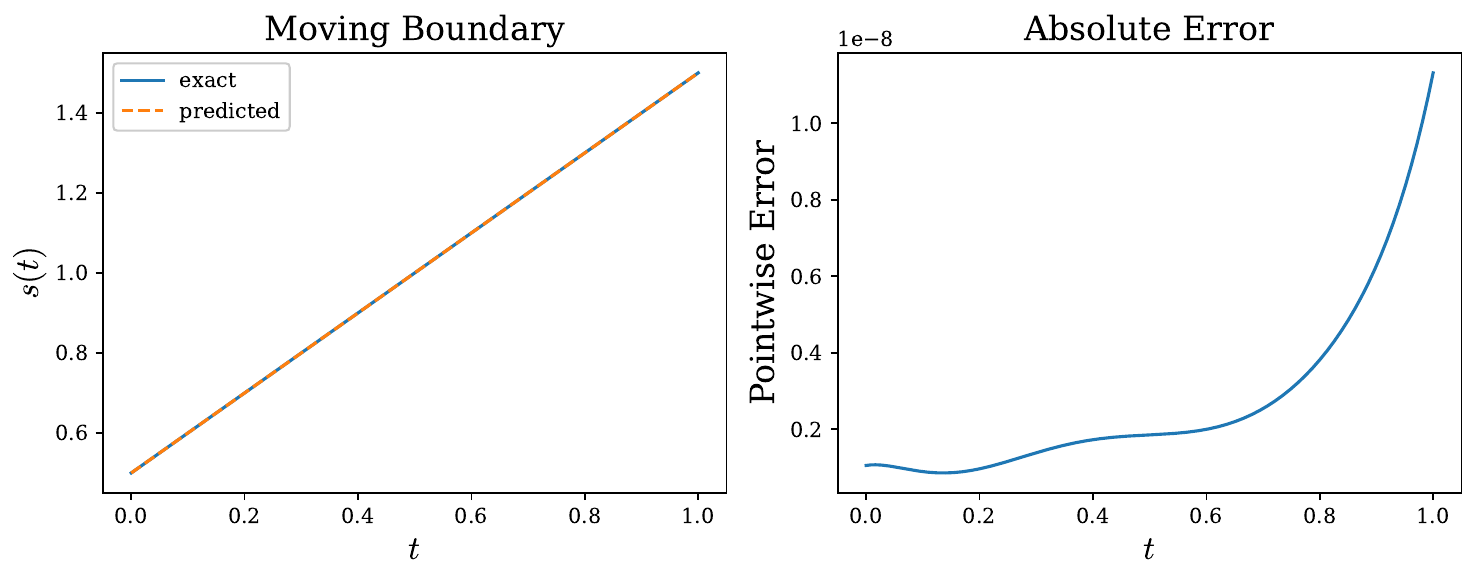}}    \caption{(a) The exact temperature \(u\), predict temperature \(\hat{u}\), and the absolute error \(|u - \hat{u}|\) for Example 1-1 using the present PINN method. (b) The moving boundary and the absolute error \(|s - \hat{s}|\) for Example 1-1.} \label{fig:ex1}
\end{figure}

Figure~\ref{fig:ex1_a} shows a comparison of the exact temperature \(u\), the predicted temperature \(\hat{u}\), and the pointwise absolute error over the entire space-time domain, with the moving interface indicated by a black line. In Figure~\ref{fig:ex1_b}, the moving interface and corresponding absolute error over the time interval $(0,1)$ are plotted. One can see that the present method accurately tracks the moving interface while providing a highly accurate prediction of the temperature.

\paragraph{Example 1-2 (inverse problem type I)}
In this test, we consider the inverse problem corresponding to Example~1-1, in which the boundary conditions are unknown, but the temperature at the final time is prescribed. This problem has also been studied in \cite{WP21}. To address it, we modify the loss function by removing the boundary condition term and including a term corresponding to the final time condition. The new loss function is defined as
\begin{align}
    \begin{split}
        \text{Loss}(\bft)
        & = \frac{1}{M_r} \sum_{i=1}^{M_r}\left|L_r(\bx^i, t^i; \bft)\right|^2
        + \frac{1}{M_0} \sum_{i=1}^{M_0}\left|L_0(\bx^i_0, 0; \bft)\right|^2 \\
        & + \frac{1}{M_T} \sum_{i=1}^{M_T}\left|L_{T}(\bx^i_{T}, T; \bft)\right|^2
        + \frac{1}{M_I} \sum_{k=1}^3 \sum_{i=1}^{M_I} \left|L_{I_k}(\bx^i_{I}, t^i_I; \bft)\right|^2,
    \end{split} \label{loss_inv1}
\end{align}
with
\begin{align}
    L_{T}(\bx_T^i, T; \bft) = \hat{u}(\bx_T^i, T; \bft) - u(\bx_T^i, T),
\end{align}
where \(T\) is the final time, and we sample \(M_T\) points for the final time condition,
$$
    \{(\bx^i_{T}, T)\}_{i=1}^{M_T}, \quad \bx^i_T\in\Omega.
$$

\begin{figure}[!ht]
    \centering
    \subfigure[]{\label{fig:ex1_inv_a}\includegraphics[width=\textwidth]{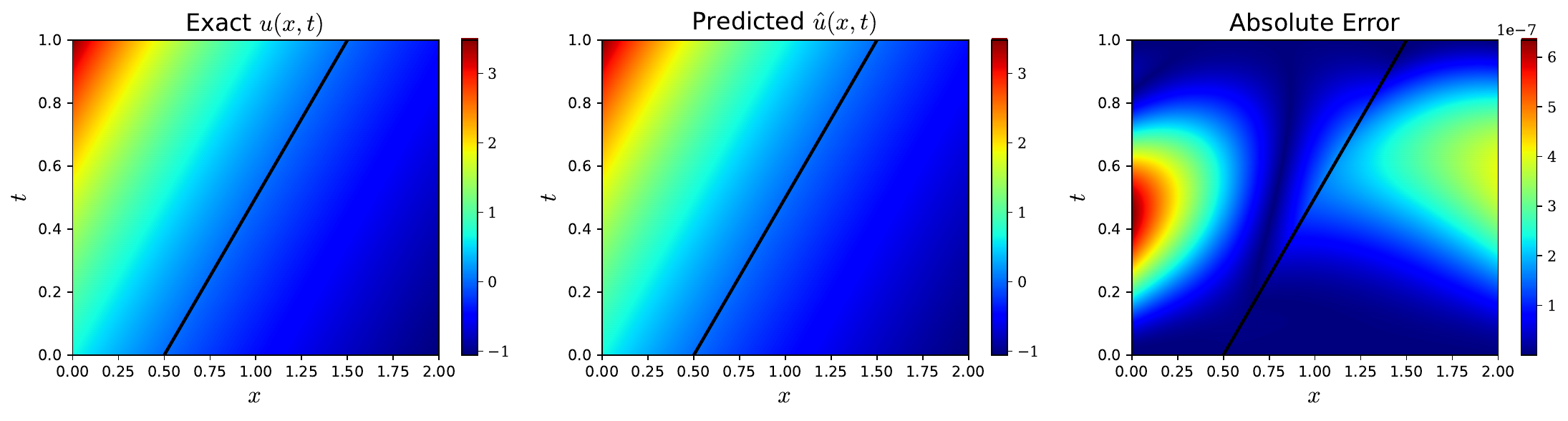}}
    \subfigure[]{\label{fig:ex1_inv_b}\includegraphics[width=.75\textwidth]{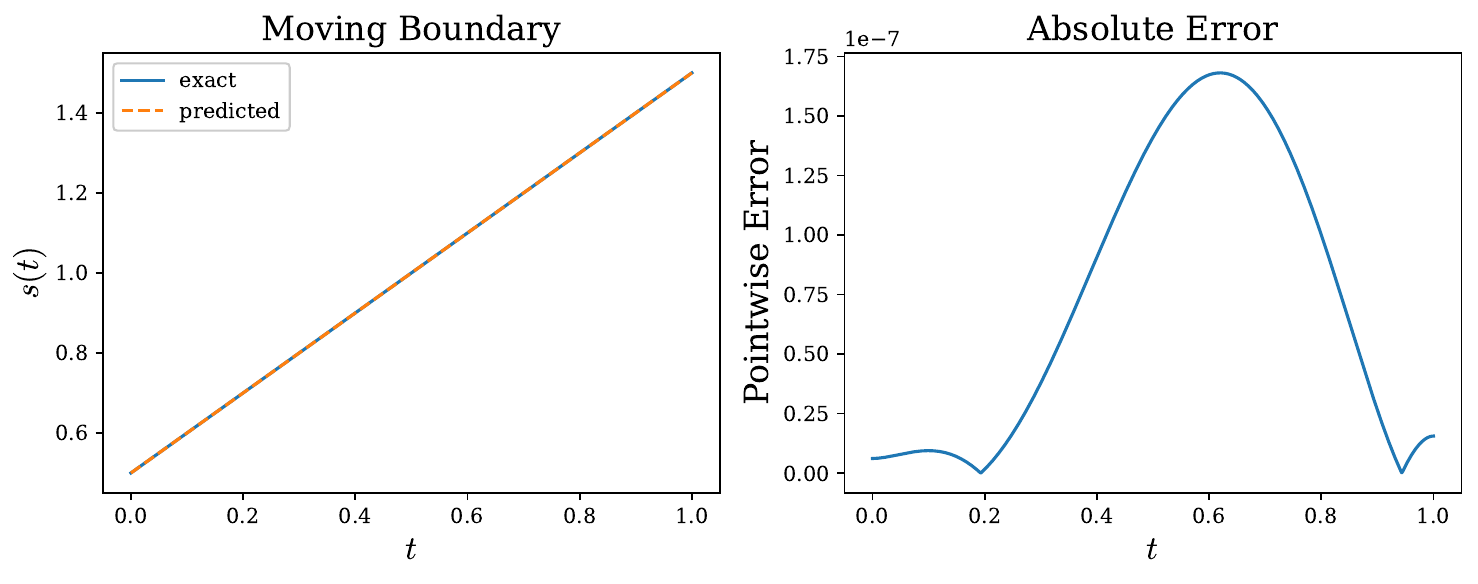}}    \caption{(a) The exact temperature \(u\), predicted temperature \(\hat{u}\), and the absolute error \(|u - \hat{u}|\) for Example 1-2 using the present PINN method. (b) The moving boundary and the absolute error \(|s - \hat{s}|\).} \label{fig:ex1_inv}
\end{figure}

The inverse problem is solved using the same training procedure as in Example 1-1. Table~\ref{tab:ex1_errors_inv} and Figure~\ref{fig:ex1_inv} present the relative \(L^2\) errors of \(\hat{u}\) and \(\hat{s}\), as well as the predicted temperature and moving boundary. We can see that the proposed method accurately solves the inverse problem, achieving a relative \(L^2\) error on the order of \(10^{-9}\).

\begin{table}[!ht]
    \begin{tabular}{c|c|cc}
        \toprule
        Method                  & Network Architectures              & $\|\hat{u}-u\|_2/\|u\|_2$ & $\|\hat{s}-s\|_2/\|s\|_2$ \\
        \hline
        Wang et al. \cite{WP21} & \makecell{u: (2, 100, 100, 100, 2)                                                         \\ s: (1, 100, 100, 100, 1)} & 1.91e-03                    & 7.01e-04                                 \\
        \hline
        Present                 & \makecell{u: (3, 32, 1)                                                                    \\ s: (1, 32, 1)}             & 8.34e-09                     & 4.68e-09              \\
        \bottomrule
    \end{tabular}
    \caption{Relative \(L^2\) errors of \(\hat{u}\) and \(\hat{s}\) for Example 1-2 inverse problem type I. The errors are averaged over ten runs. The values in Wang et al.~\cite{WP21} are taken directly from the paper.}
    \label{tab:ex1_errors_inv}
\end{table}

\paragraph{Example 1-3 (inverse problem type II)}

\begin{figure}[!ht]
    \centering
    \subfigure[]{\label{fig:ex1_inv2_a}\includegraphics[width=\textwidth]{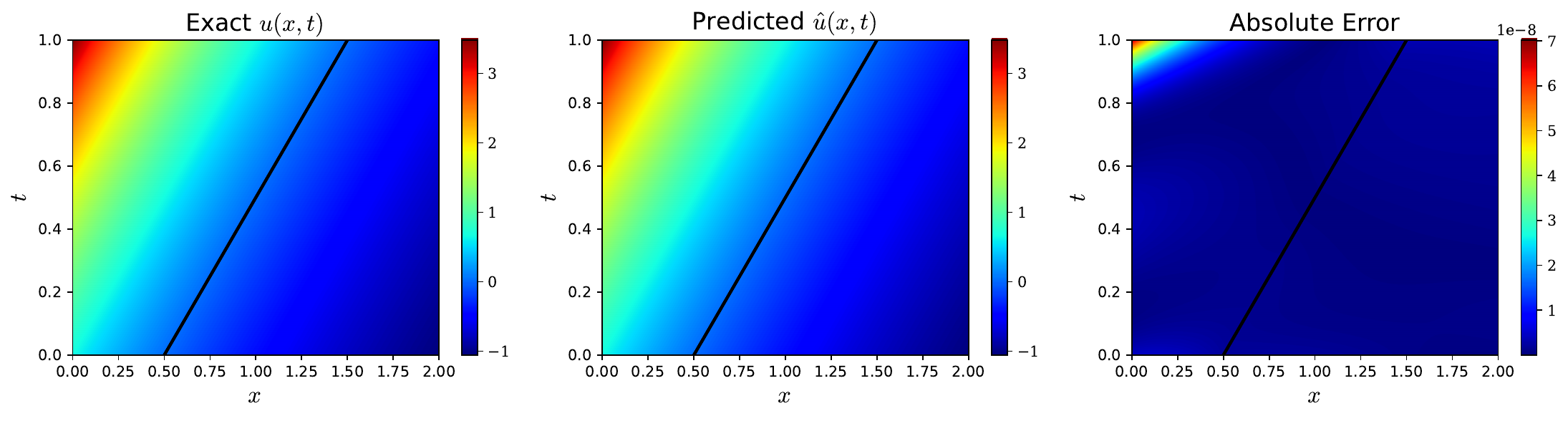}}
    \subfigure[]{\label{fig:ex1_inv2_b}\includegraphics[width=.75\textwidth]{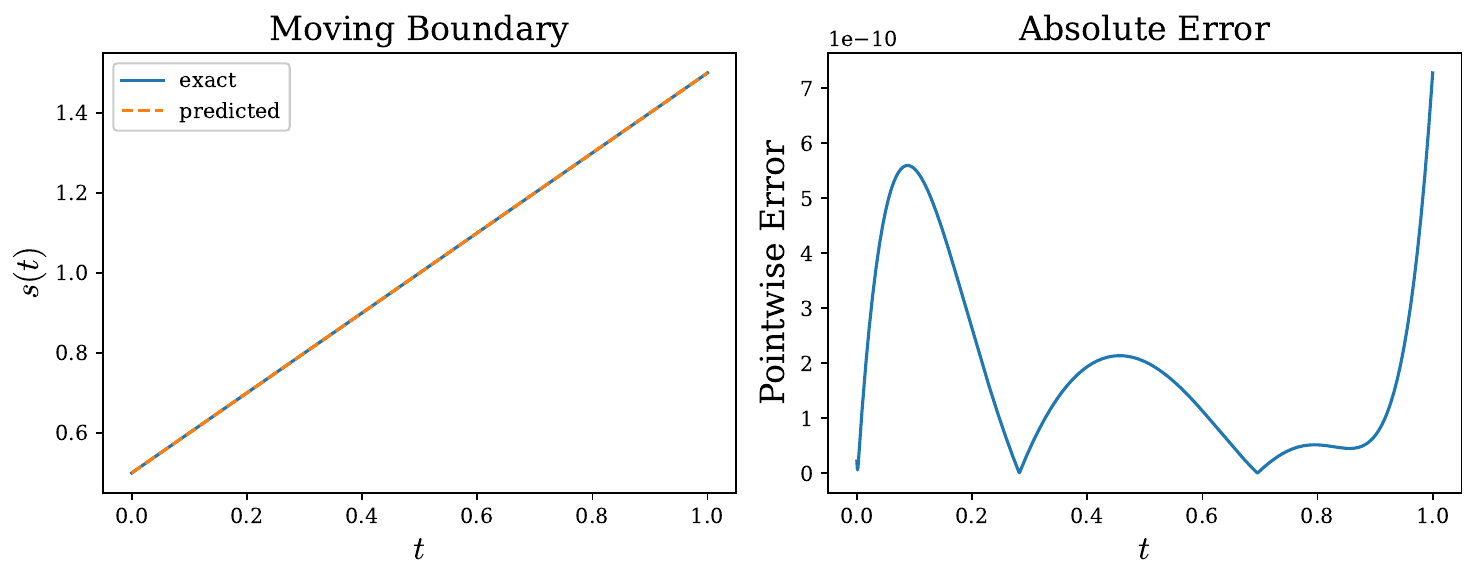}}    \caption{(a) The exact temperature \(u\), predicted temperature \(\hat{u}\), and the absolute error \(|u - \hat{u}|\) for Example 1-3 using the present PINN method. (b) The moving boundary and the absolute error \(|s - \hat{s}|\).} \label{fig:ex1_inv_2}
\end{figure}

In this test, we consider another type of inverse problem based on Example 1-1, in which the initial and boundary conditions are unknown, while some temperature measurements are known. This problem has also been considered in \cite{WP21}. Using the same training procedure as in Example 1-1, the problem can be solved by appropriately modifying the loss function as
\begin{align}
    \begin{split}
        \text{Loss}(\bft)
        & = \frac{1}{M_r} \sum_{i=1}^{M_r}\left|L_r(\bx^i, t^i; \bft)\right|^2
        + \frac{1}{M_I} \sum_{k=1}^3 \sum_{i=1}^{M_I} \left|L_{I_k}(\bx^i_{I}, t^i_I; \bft)\right|^2 \\
        &+ \frac{1}{M_{\text{data}}} \sum_{i=1}^{M_{\text{data}}} \left|L_{\text{data}}(\bx^i_{\text{data}}, t^i_{\text{data}}; \bft)\right|^2,
    \end{split} \label{loss_inv2}
\end{align}
with
\begin{align}
    L_{\text{data}}(\bx_{\text{data}}^i, t_{\text{data}}; \bft) = \hat{u}(\bx_{\text{data}}^i, t_{\text{data}}; \bft) - u(\bx_{\text{data}}^i, t^i_{\text{data}}),
\end{align}
where \(M_{\text{data}}\) is the number of points for the prescribed temperature measurements, and \(\{(\bx^i_{\text{data}}, t^i_{\text{data}})\}_{i=1}^{M_{\text{data}}}\) are the data points. We take \(M_{\text{data}} = 20\) and randomly sample data points from the space-time domain. Table~\ref{tab:ex1_errors_inv_2} and Figure~\ref{fig:ex1_inv_2} summarize the relative \(L^2\) errors of \(u\) and \(s\), and the predicted temperature and moving boundary. The proposed method accurately solves the inverse problem, achieving a relative \(L^2\) error on the order of \(10^{-9}\).

\begin{table}[!ht]
    \begin{tabular}{c|c|cc}
        \toprule
        Method                  & Network Architectures              & $\|\hat{u}-u\|_2/\|u\|_2$ & $\|\hat{s}-s\|_2/\|s\|_2$ \\
        \hline
        Wang et al. \cite{WP21} & \makecell{u: (2, 100, 100, 100, 2)                                                         \\ s: (1, 100, 100, 100, 1)} & 2.57e-03                    & 3.93e-04                                 \\
        \hline
        Present                 & \makecell{u: (3, 32, 1)                                                                    \\ s: (1, 32, 1)}             & 2.89e-09                     & 2.55e-09             \\
        \bottomrule
    \end{tabular}
    \caption{Relative \(L^2\) errors of \(\hat{u}\) and \(\hat{s}\) for Example 1-3. The errors are averaged across ten runs. The values in Wang et al.~\cite{WP21} are taken directly from the paper.}
    \label{tab:ex1_errors_inv_2}
\end{table}

In addition, we perform a sensitivity study similar to the approach in \cite{WP21}. We add different levels of Gaussian noise with varying standard deviation \(\sigma\) to the temperature measurements, and the number of data points is also varied. Table~\ref{tab:ex1_sensitivity_u} and Table~\ref{tab:ex1_sensitivity_s} report the relative \(L^2\) errors of \(\hat{u}\) and \(\hat{s}\) under different noise levels and numbers of measurements. The reported errors are averaged over five runs. The results indicate that our proposed methodology accurately captures both the temperature and the moving interface, even in the presence of additive noise.

\begin{table}[!ht]
    \centering
    \begin{tabular}{r|rrrr}
        \toprule
        \(M_{\text{data}}\) & \(\sigma=0.01\) & \(\sigma=0.02\) & \(\sigma=0.05\) & \(\sigma=0.1\) \\
        \midrule
        20                  & 4.5501e-03      & 9.9876e-03      & 1.8851e-02      & 4.2803e-02     \\
        50                  & 2.0612e-03      & 5.6282e-03      & 1.3883e-02      & 3.7048e-02     \\
        100                 & 2.2317e-03      & 5.3921e-03      & 1.1385e-02      & 1.6072e-02     \\
        \bottomrule
    \end{tabular}
    \caption{Relative \(L^2\) errors of the temperature \(\hat{u}\) for Example 1-3 inverse problem type II with different levels of noise and number of data points. The errors are averaged across five runs.}\label{tab:ex1_sensitivity_u}
\end{table}
\begin{table}
    \centering
    \begin{tabular}{r|rrrr}
        \toprule
        \(M_{\text{data}}\) & \(\sigma=0.01\) & \(\sigma=0.02\) & \(\sigma=0.05\) & \(\sigma=0.1\) \\
        \midrule
        20                  & 1.4921e-03      & 5.2383e-03      & 1.0077e-02      & 1.8047e-02     \\
        50                  & 9.8147e-04      & 2.2767e-03      & 5.4934e-03      & 1.4523e-02     \\
        100                 & 2.2214e-03      & 2.7569e-03      & 3.6821e-03      & 1.2022e-02     \\
        \bottomrule
    \end{tabular}
    \caption{Relative \(L^2\) errors of the interface \(\hat{s}\) for Example 1-3 inverse problem type II with different levels of noise and number of data points. The errors are averaged across five runs.}\label{tab:ex1_sensitivity_s}
\end{table}

\paragraph{Example 2-1}

The second test is modified from the example presented in Section~3.3.1 of \cite{WP21}, which considers a two-dimensional one-phase Stefan problem. Here, we extend the problem to the two-phase case.

We set \(\Omega = (0, 1) \times (0, 2)\), \(t\in (0, 1)\), \(k^+=1\), \(k^- = 1/2\), and \(\beta = 3/2\). The exact solution is given by
\[
    u(x,y,t) =
    \begin{cases}
        e^{-(\frac{5}{4} t - x + \frac{y}{2} + \frac{1}{8})} -1, & x \leq \frac{1}{2} y + \frac{5}{4} t + \frac{1}{8}, \\
        e^{(\frac{5}{4} t - x + \frac{y}{2} + \frac{1}{8})} -1,  & x >  \frac{1}{2} y + \frac{5}{4} t + \frac{1}{8},
    \end{cases}
\]
so the moving interface $s(y,t) = \frac{1}{2} y + \frac{5}{4} t + \frac{1}{8}$. One can verify that the melting temperature condition Eq.~(\ref{freezing}) and Stefan condition Eq.~(\ref{moving}) are all satisfied at the moving interface. As in the previous examples, the initial and boundary conditions are prescribed using the exact solution, and Dirichlet boundary conditions are imposed along the domain boundaries.

We choose \((M_r, \tilde{M}_b) = (4096, 2048)\) training points and train the network using the LM optimizer for a maximum of $2000$ iterations. Both \(U_{\theta_1}\) and \(\hat{s}_{\theta_2}\) use only one hidden layer network  but with different input dimensions. Table~\ref{tab:ex2_errors} reports the $L^\infty$ errors of \(u\) and \(s\), as well as the testing losses for different number of neurons used in the hidden layer and training points.    The $L^\infty$ errors are computed by sampling \(M_r = 10^6\) and \(\tilde{M}_b = 8\times 10^6\) points and are averaged over five runs. As shown in the table, increasing the number of neurons in the hidden layer reduces the $L^\infty$ errors of \(\hat{u}\), \(\hat{s}\), as well as the testing loss, up to a certain point beyond which further improvement saturates.

\begin{table}[!ht]
    \begin{center}
        \begin{tabular}{c|c|ccc}
            \toprule
            $(M_r, \tilde{M}_b)$ & \# of neurons (hidden layer) & $\|\hat{u}-u\|_\infty$ & $\|\hat{s}-s\|_\infty$ & $\Loss(\bft)$ \\
            \midrule
                                 & 4                            & 1.5816e-07             & 1.1564e-07             & 9.8332e-14    \\
            (512, 512)           & 8                            & 4.3673e-09             & 2.5887e-09             & 1.3795e-16    \\
                                 & 16                           & 1.3621e-08             & 5.2354e-09             & 1.4827e-15    \\
                                 & 32                           & 5.9526e-09             & 4.0204e-09             & 1.5763e-16    \\
            \midrule
                                 & 4                            & 1.8566e-05             & 9.3263e-07             & 6.9202e-09    \\
            (4096, 2048)         & 8                            & 4.4402e-09             & 2.5239e-09             & 7.6167e-17    \\
                                 & 16                           & 3.8251e-09             & 3.6937e-10             & 6.7233e-17    \\
                                 & 32                           & 6.1651e-09             & 4.5470e-09             & 7.7621e-17    \\
            \bottomrule
        \end{tabular}
        \caption{$L^\infty$ errors of $\hat{u}$, $\hat{s}$, and the testing loss for Example 2-1.} \label{tab:ex2_errors}
    \end{center}
\end{table}

\paragraph{Example 2-2 (inverse problems)}

We also consider the inverse problems corresponding to Example 2-1. The loss functions for the type I and type II inverse problems are modified in the same manner as in Example~1-2 and Example~1-3, respectively. Table~\ref{tab:ex2_errors_inv} reports the $L^\infty$ errors of \(\hat{u}\), \(\hat{s}\), and the testing losses for different number of neurons used in the hidden layer and  training points for both types of inverse problems.  For the type II inverse problem, $M_{data}=100$ temperature measurements are used. The results indicate that the proposed method can accurately solve both types of inverse problems.

\begin{table}
    \centering
    \begin{tabular}{c|c|ccc}
        \toprule
        Problem    & \# of neurons (hidden layer) & $\|\hat{u}-u\|_\infty$ & $\|\hat{s}-s\|_\infty$ & $\Loss(\bft)$ \\
        \midrule
        Inverse I  & 32                           & 6.123e-09              & 2.806e-09              & 6.540e-17     \\
        Inverse II & 32                           & 1.374e-08              & 6.460e-09              & 6.827e-17     \\
        \bottomrule
    \end{tabular}
    \caption{$L^\infty$ errors of \(\hat{u}\), \(\hat{s}\), and the testing loss for Example 2-2 inverse problems. The errors are averaged over ten runs.}
    \label{tab:ex2_errors_inv}
\end{table}

\paragraph{Example 3}

In this example, we investigate a physical scenario similar to ice-water solidification problem  where the exact solution is not available. The computational domain is taken to be \(\Omega = (0, 1) \times (0, 1)\), \(t\in (0, 1)\), and the Stefan number \(\beta = 1\). The initial interface is  given by
\[
    x = s_0(y) = 1 / 4 + \varepsilon \cos(2 \pi  y),
\]
and the initial temperature is prescribed as
\[
    u_0(x,y) = \begin{cases}
        (-1 + x  y^2
        (y - 1)^2)  (1 / 4 + \varepsilon  \cos(2 \pi  y) - x)  & \text { if } x \leq s_0(y), \\
        (1 - (1 - x)  y^2
        (y - 1)^2)  (x - (1 / 4 + \varepsilon  \cos(2 \pi y))) & \text { if } x > s_0(y),
    \end{cases}
\]
where \(\varepsilon=0.05\). It is worthy noting that this initial interface $x = s_0(y)$ and initial temperature $u_0(x,y)$ are carefully designed so that the freezing temperature is exactly zero along the interface $x = s_0(y)$. Furthermore, $u_0(x,y)$ satisfies zero Neumann boundary condition  at $y=0$ and $y=1$ indicating that both boundaries of $y$ are insulated (no heat flux across the boundaries of $y$). This can be easily checked by taking the derivative with respect to $y$ for $u_0(x,y)$ and evaluating the derivative at $y=0$ and $y=1$ accordingly. Therefore, we can enforce zero Neumann boundary condition along the \(y\)-direction for $u(x,y,t)$ naturally. On the other hand, we use the Dirichlet bounday condition of the initial temperature $u_0(x,y)$ for the boundary at $x=0$ and $x=1$ so that $u(0,y,t)=-(1 / 4 + \varepsilon  \cos(2 \pi  y))$ and $u(1,y,t)=1 - (1 / 4 + \varepsilon  \cos(2 \pi  y))$, respectively. In this setup, one can regard $x < s(y,t)$ as the ice region, while $x > s(y,t)$ as the water region.

We fix the thermal diffusivity $k^+$ to be \(k^+ = 0.1\), while vary two different \(k^-=1, 3\) to see how they affect the interface moving via the Stefan condition Eq.~(\ref{moving}). The network is trained using the LM optimizer for $2000$ iterations, with the hidden-layer architecture \([16, 16, 16, 16]\) for both \(U_{\theta_1}\) and \(\hat{s}_{\theta_2}\), i.e., $4$ hidden layers and $16$ neurons at each hidden layer. We use \(M_r = 2^{16}\) and \(\tilde{M}_b = 2^{17}\) training points and sample \(\tilde{M}_b = 8\times 10^6\) and \(M_r = 10^6\) points for testing. Since the exact solution is unavailable, only the testing losses are evaluated. The final testing loss is  \(5.68 \times 10^{-4}\) for \(k^- = 1\) and \(4.22 \times 10^{-4}\) for \(k^- = 3\).

\begin{figure}[!ht]
    \centering
    \includegraphics[width=.9\textwidth]{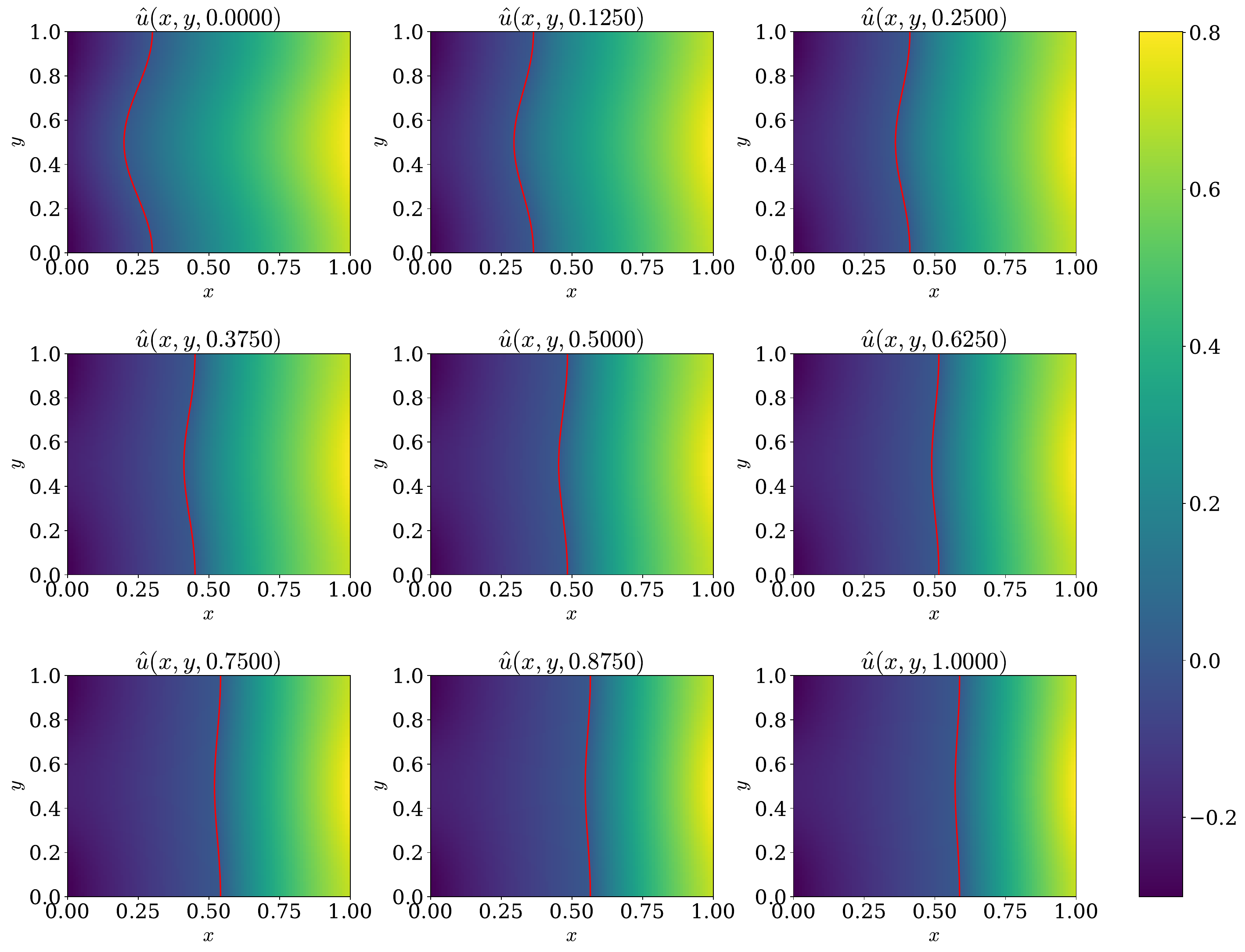}
    \caption[]{The temperature \(\hat{u}(x,y,t)\) at different time instances for Example~3 with \(k^-=1\). The red line indicates the position of the interface \(x=\hat{s}(y,t)\).}\label{fig:ex3_2}
\end{figure}

\begin{figure}[!ht]
    \centering
    \includegraphics[width=.9\textwidth]{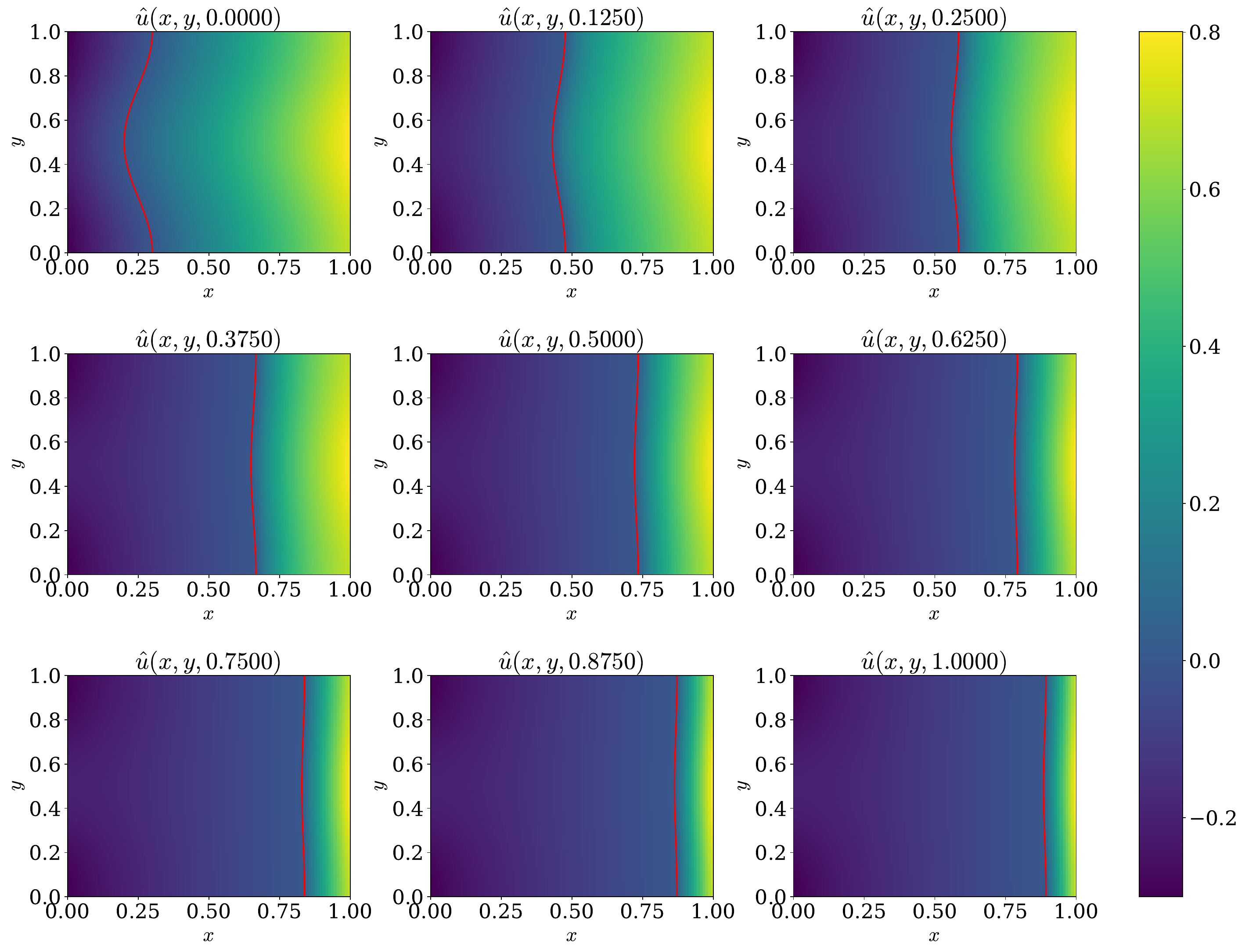}
    \caption[]{The temperature \(\hat{u}(x,y,t)\) at different time instances for Example~3 with \(k^-=3\). The red line indicates the position of the interface \(x=\hat{s}(y,t)\).}\label{fig:ex3}
\end{figure}

\begin{figure}[!ht]
    \centering
    \includegraphics[width=.9\textwidth]{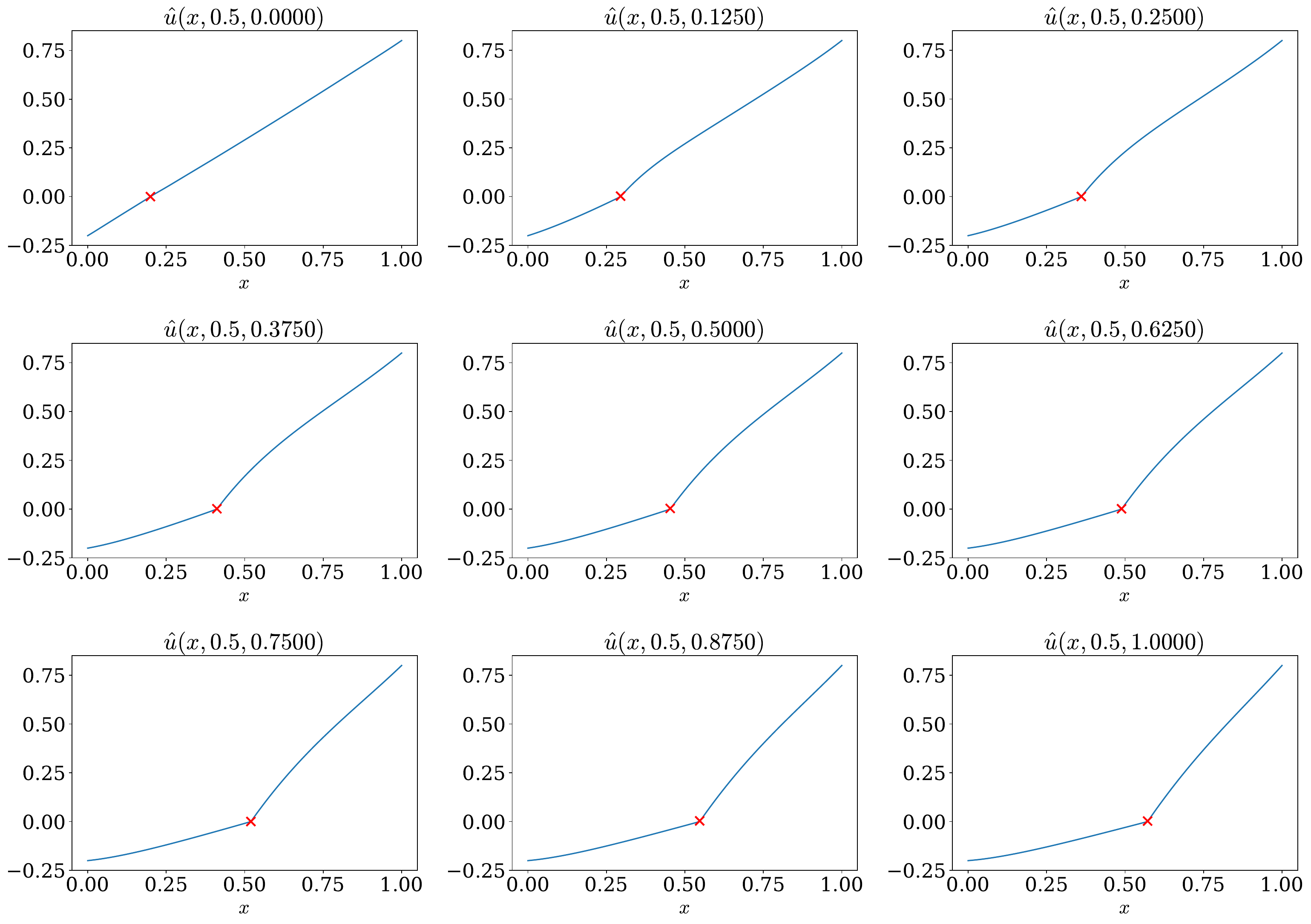}
    \caption[]{Cross-section of the temperature \(\hat{u}(x,y,t)\) at $y=0.5$ at different time instances for Example~3 with \(k^-=1\). The red cross indicates the position of the interface \(x=\hat{s}(0.5,t)\).}\label{fig:ex3_section_2}
\end{figure}

\begin{figure}[!ht]
    \centering
    \includegraphics[width=.9\textwidth]{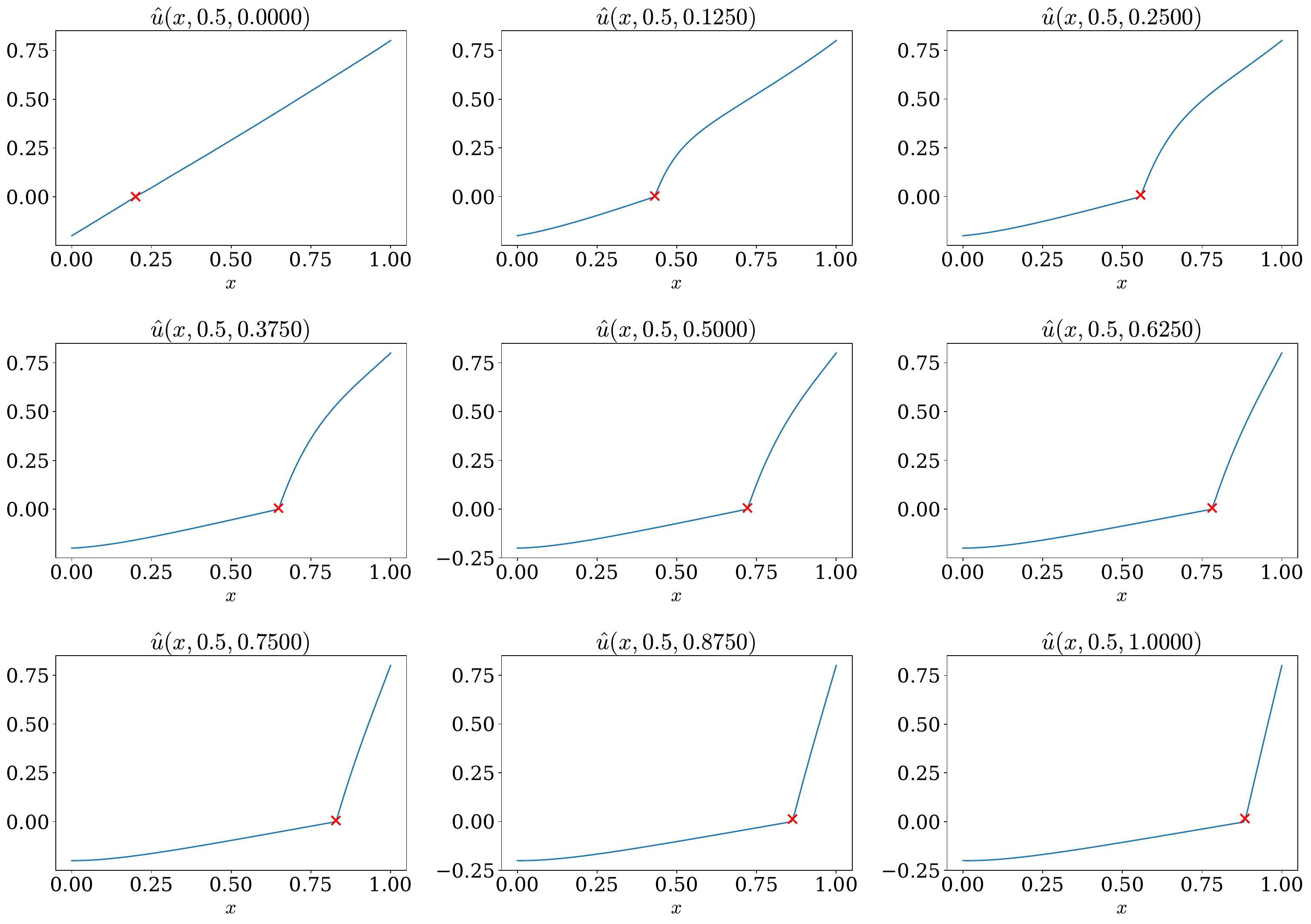}
    \caption[]{Cross-section of the temperature \(\hat{u}(x,y,t)\) at $y=0.5$ at different time instances for Example~3 with \(k^-=3\). The red cross indicates the position of the interface \(x=\hat{s}(0.5,t)\). }\label{fig:ex3_section}
\end{figure}

When the thermal diffusivity $k^+$ is fixed, increasing $k^-$ makes the magnitude of the thermal diffusivity jump larger. This increased jump leads to a greater net heat flux accumulation at the interface, which, according to the Stefan condition Eq.~(\ref{moving}), corresponds to a higher interface normal velocity. Figures~\ref{fig:ex3_2} and \ref{fig:ex3} show the temperature and the interface at different times of testing results for \(k^-=1\) and \(k^-=3\), respectively. One can immediately see that the interface moves faster when \(k^-=3\) compared to \(k^-=1\), which aligns with physical expectations. Furthermore, as time evolves, the interface moves slower and its shape tends towards a more vertical configuration due to the influence of thermal diffusion. Eventually, both the temperature distribution and the interface position will asymptotically reach a steady state. Figures~\ref{fig:ex3_section_2} and \ref{fig:ex3_section} show the corresponding temperature fields at cross-section $y=0.5$ for \(k^-=1\) and \(k^-=3\), respectively. The results clearly show that the numerical method exactly resolves the cusp-discontinuity in the temperature profile at the interface. Moreover, the temperature distribution tends toward a more linear profile, validating the conclusion that the system is asymptotically approaching a steady state.

\paragraph{Example 4}

The fourth example is the Mullins-Sekerka instability problem, first introduced by Mullins and Sekerka~\cite{MS64} and later studied in greater detail by Almgren~\cite{Almgren93} and Strain~\cite{Strain88}. It describes a physical phenomenon in which small perturbations to a flat interface grow arbitrarily large in amplitude. This example corresponds to a two-dimensional, two-phase Stefan problem with an unstable moving interface. The objective of this example is to examine the ability of the present method to capture the dynamics of the unstable interface.

The computational domain is taken to be \(\Omega = (0,1) \times (0, \pi)\), \(t \in (0.3, 0.8)\), \(k^-=1\), \(k^+ = 1\), and the Stefan number \(\beta = 1\). The initial interface is given by
\[
    x = s_0(y) = Vt_0 + \varepsilon \cos(Ky),
\]
and the initial temperature is prescribed as
\[
    u_0(x,y) = \begin{cases}
        0,                                             & \text { if } x \leq s_0(y), \\
        -1 + e^{-V(x - (Vt_0 + \varepsilon\cos(Ky)))}, & \text { if } x > s_0(y),
    \end{cases}
\]
where \(V=1\), \(K=5\), and \(\varepsilon = 0.05\). Here, to be consistent, we use the same Neumann boundary conditions for $u(x,y,t)$ as the ones obtained from the initial temperature $u_0(x,y)$.

\begin{figure}[!ht]
    \centering
    \includegraphics[width=.9\textwidth]{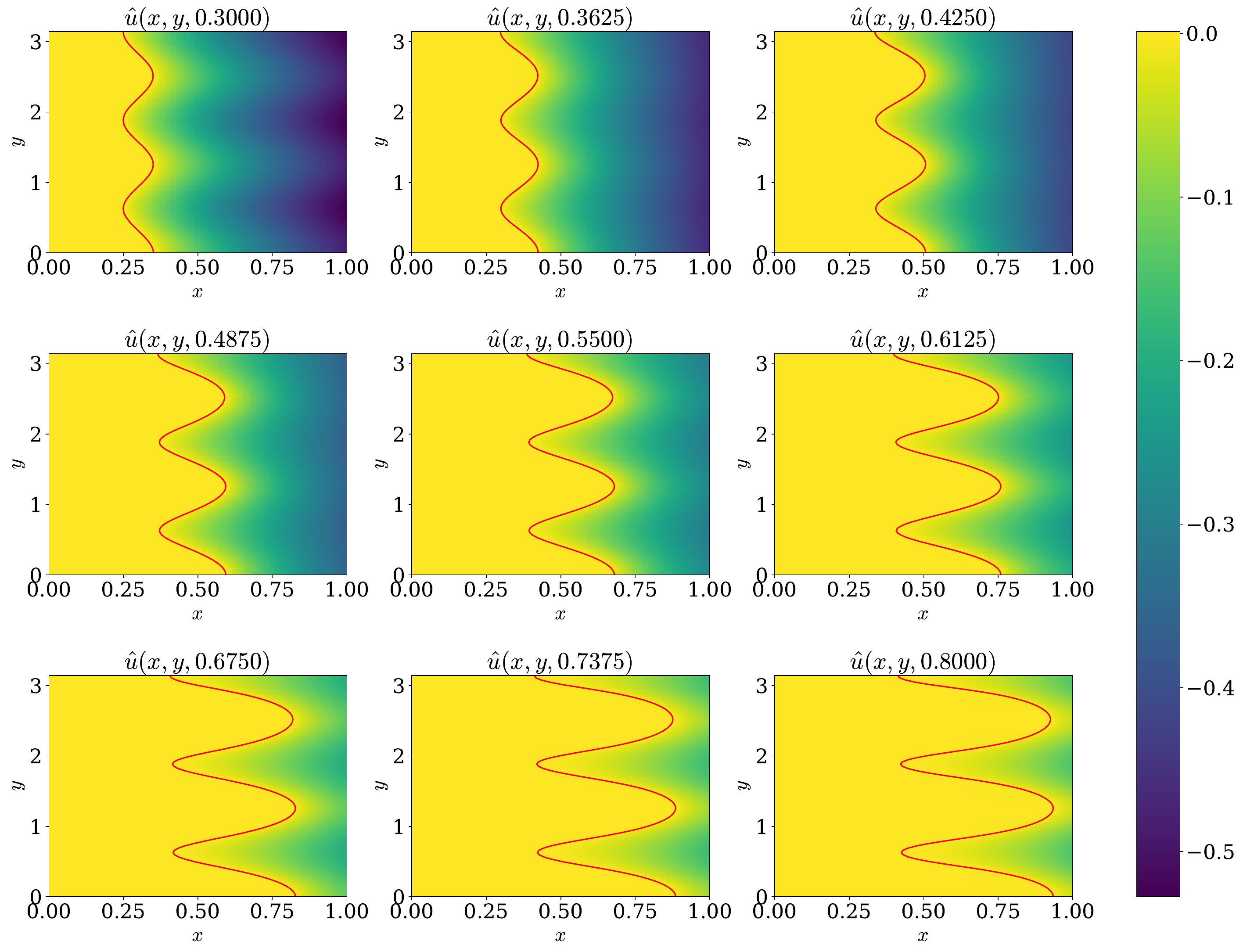}
    \caption[]{\(\hat{u}(x,y,t)\) at different time instances \(t\) for Example 4. The red line indicates the interface \(x=\hat{s}(y,t)\).}\label{fig:ex4}
\end{figure}

\begin{figure}[!ht]
    \centering
    \includegraphics[width=.95\textwidth]{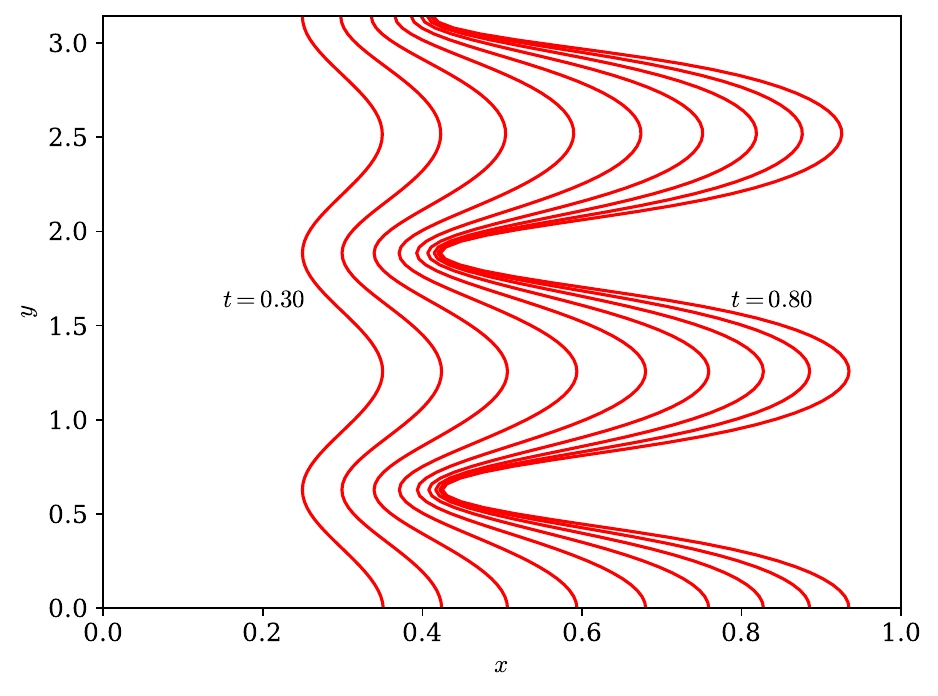}
    \caption{Interface position \(x=\hat{s}(y,t)\) at different time instances for Example 4. The interface grows from left to right as
        time evolves.}
    \label{fig:ex4_interface}
\end{figure}

The training procedure and network architecture are the same as those used in Example~3. With the same testing evaluation process, we obtain a final testing loss of \(1.33 \times 10^{-5}\). Figure~\ref{fig:ex4} shows the temperature and the interface at several time instances. To be more clearly, the time evolution of the interface is presented separately in Figure~\ref{fig:ex4_interface}. Unlike the previous experiment, Example~3, where the interface amplitude was damped, here the amplitude of the interface tends to grow over time. Thus, it is evident that our methodology can capture the dynamics of Mullins-Sekerka instability behavior qualitatively.

\section{Conclusion}
\label{sec:conclusion}

In this paper, we proposed a cusp-capturing PINN method for solving the two-dimensional two-phase Stefan problem. Building upon the physics-informed neural network framework, our approach incorporates a cusp-capturing technique that introduces an auxiliary level-set-type variable to a neural network to implicitly represent the interface and incorporate jump conditions directly into the loss function. This formulation allows simultaneous learning of both the temperature field and the interface evolution in a unified and mesh-free manner.

We demonstrated the effectiveness of the proposed methodology in accurately capturing both the moving interface and the temperature field. The results show that the cusp-capturing PINNs can be successfully applied to a variety of Stefan problems, including forward and inverse formulations, as well as the Mullins-Sekerka instability problem. Our approach demonstrates superior accuracy and robustness compared to previous PINN-based methods. Moreover, we have shown that the Levenberg-Marquardt optimizer serves as a stable training strategy for the cusp-capturing PINNs. As future work, we plan to extend this framework to more complex Stefan problems and explore the integration of the cusp-capturing technique into other classes of physics-informed neural network models.

\section*{Acknowledgments}

T.-S. Lin and M.-C. Lai acknowledge the supports by National Science and Technology Council, Taiwan, under research grants 111-2628-M-A49-008-MY4 and 114-2124-M-390-001. M.-C. Lai acknowledge the support by National Science and Technology Council, Taiwan, under research grant 113-2115-M-A49-014-MY3. T.-S. Lin also acknowledge the supports by National Center for Theoretical Sciences, Taiwan.

\end{document}